\documentclass[dvipsnames,unnumsec,webpdf,contemporary,large,numbers]{oup-authoring-template_edit}

\usepackage[T1]{fontenc}
\usepackage{microtype}
\usepackage{bm}
\usepackage{nicefrac}

\newcommand{\graphic}[2]{\includegraphics[width=#2\linewidth, type=pdf,ext=.pdf,read=.pdf]{#1}}

\newcommand{\dee}{\mathrm{d}}
\newcommand{\lcdm}{\Lambda\mathrm{CDM}}
\newcommand{\mpc}{\mathrm{Mpc}}

\newcommand{\impc}{\mpc^{-1}}
\newcommand{\om}{\Omega_\mathrm{m}}

\begin{document}

\journaltitle{Journal Title Here}
\DOI{DOI HERE}
\copyrightyear{2022}
\pubyear{2022}
\appnotes{~}

\firstpage{1}

\title[Simple lessons from complex learning]{Simple lessons from complex learning: what a neural network model learns about cosmic structure formation}

\author[1,2,$\ast$]{Drew Jamieson}
\author[3,4,$\ast$]{Yin Li}
\author[3,5]{Siyu He}
\author[3,7]{Francisco Villaescusa-Navarro}
\author[3,7]{Shirley Ho}
\author[3,6]{Renan Alves de Oliveira}
\author[8]{David N. Spergel}

\authormark{Jamieson \& Li et al.}

\address[1]{Max-Planck-Institut f\"ur Astrophysik, Karl-Schwarzschild-Straße 1, 85748 Garching, Germany}
\address[2]{Department of Physics and Astronomy, Stony Brook University,
Stony Brook, NY 11794-3800, USA}
\address[3]{Center for Computational Astrophysics,
Flatiron Institute, 162 5th Avenue, New York, NY 10010, USA}
\address[4]{Center for Computational Mathematics,
Flatiron Institute, 162 5th Avenue, New York, NY 10010, USA}
\address[5]{Department of Physics, Carnegie Mellon University,
Pittsburgh, PA 15213, USA}
\address[6]{Centro de Ciências Exatas, Universidade Federal do Espírito
Santo. Av. Fernando Ferrari, 514. 29075-910. Vitória, ES, Brazil}
\address[7]{Department of Astrophysical Sciences, 4 Ivy Lane, Princeton
University, Princeton, NJ 08544, USA}
\address[8]{Simons Foundation, 160 5th Avenue, New York, NY 10010, USA}

\corresp[$\ast$]{Corresponding authors.
\href{mailto:jamieson@mpa-garching.mpg.de}{jamieson@mpa-garching.mpg.de}
\& \href{mailto:eelregit@gmail.com}{eelregit@gmail.com}}


\abstract{
    We train a neural network model to predict the full phase space evolution of cosmological N-body simulations. Its success implies that the neural network model is accurately approximating the Green's function expansion that relates the initial conditions of the simulations to its outcome at later times in the deeply nonlinear regime. We test the accuracy of this approximation by assessing its performance on well understood simple cases that have either known exact solutions or well understood expansions. These scenarios include spherical configurations, isolated plane waves, and two interacting plane waves: initial conditions that are very different from the Gaussian random fields used for training. We find our model generalizes well to these well understood scenarios, demonstrating that the networks have inferred general physical principles and learned the nonlinear mode couplings from the complex, random Gaussian training data. These tests also provide a useful diagnostic for finding the model's strengths and weaknesses, and identifying strategies for model improvement. We also test the model on initial conditions that contain only transverse modes, a family of modes that differ not only in their phases but also in their evolution from the longitudinal growing modes used in the training set. When the network encounters these initial conditions that are orthogonal to the training set, the model fails completely. In addition to these simple configurations, we evaluate the model's predictions for the density, displacement, and momentum power spectra with standard initial conditions for N-body simulations. We compare these summary statistics against N-body results and an approximate, fast simulation method called COLA. Our model achieves percent level accuracy at nonlinear scales of $k\sim 1\ \impc\, h$, representing a significant improvement over COLA.
}


\maketitle

\section{Significance Statement}

    The success of the neural network in approximating the nonlinear dynamics of structure growth in cosmology will enable future cosmological analyses. Our neural network model is able to predict the outcome of cosmological structure formation robustly, accurately, and efficiently at the field level for the full dark matter phase space in the deeply nonlinear regime. Our probing of the properties of this network reveal that the network has learned to approximate the Green's function expansion of the structure growth (equation 8). While the network is trained on random Gaussian initial conditions, it accurately describes the evolution of well understood systems with highly correlated initial phases and highly non-Gaussian initial conditions. Its successes and failures yields more general insights into the properties of the neural network.
        
\section{Introduction}
	\label{sec:intro}
	
	Physicists tend to gain understanding of various systems by building toy models based on simplifying assumptions, which make them analytically tractable or numerically feasible.  The simplicity of such models comes with the added benefit of making them easy to interpret, since they typically contain only a small number of meaningful parameters. As alleged by Philip Anderson, ``a simplified model throws more light on the real workings of nature than any number of `ab initio' calculations of individual situations, which even where correct often contain so much detail as to conceal rather than reveal reality'' \cite{anderson1978}. However, the cost of these simplified models is their limited range of applicability, and the difficulty of generalizing them to describe a wider range of phenomena. The advent of high performance computing has made it possible to model physical systems without such simplifying assumptions and conservative parameterizations. The deep learning \citep{lecun2015deep} approach is to dramatically overparameterize a system and force the model to learn complex patterns and features of a rich data set in an efficient, automated way. The great advantage of this new modeling paradigm is the wide range of applicability and the diversity of phenomena that can be described by such models. The cost is the loss of interpretability of the model and its parameters. In this way, the principles underlying deep learning and artificial neural networks (NNs) are both antithetical and complementary to those underlying the traditional approaches of physicists. Instead of building simplified and abstract models based on physical insight, NNs learn by finding patterns in experimental, observational, and simulated data, and do so most efficiently when the patterns in the data are rich and complex rather than simple.
	
	Recently NNs have also proved useful for accelerating computationally expensive tasks by directly learning to approximate the mapping between an algorithm's input and output \cite{Pang:2016vdc,Hezaveh:2017sht,He:2018ggn,han2018,Cranmer:2019eaq,raissi2019physics,AlvesdeOliveira:2020yix,Kochkov_2021}. These applications require robust methods of testing a model's accuracy, to be confident in its performance on new data. In this work we use a set of idealized physical scenarios where exact or perturbative calculations can be performed explicitly in order to validate a NN model trained on general, complex data sets.
    
    The application we present here is in the context of structure formation in the Universe, which is a nonlinear problem involving the complex pattern formation of what is known as the cosmic web. Cosmological N-body simulations provide the most accurate model of this nonlinear process. Deriving robust and optimal parameter constraints from observations of cosmic structure requires running large numbers of these expensive simulations. To alleviate this computational cost, and to gain insight into the complicated mapping between the initial conditions and final outcomes of structure formation, simplified models such as spherical collapse \cite{Bernardeau:2001qr}, excursion set \cite{Bond:1990iw,Sheth:1999su,Zentner:2006vw}, and halo models \cite{White:2000hh,Cooray:2002dia} are often employed. These models provide intuition and qualitative predictions for cosmic structure formation that lack accuracy, whereas optimal extraction of cosmological information from observations calls for precise modeling. In recent years, NN techniques have been developed to approximate these simulations both accurately and efficiently \cite{He:2018ggn,AlvesdeOliveira:2020yix}.
    
    In this work we develop the first NN model to predict the full N-body phase space of dark matter structure formation, including displacements and velocities in the deeply nonlinear regime. We demonstrate the accuracy of our NN model by comparing its predictions against both the full N-body evolution that it was trained to reproduce, and a fast, approximate algorithm called COLA. We then valid our NN model within the context of three well understood systems: spherically symmetric configurations, isolated plane waves, and two coupled plane waves. The training data for the NN model is a set of Gaussian random fields, so these much simpler testing scenarios, which are far from Gaussian, represent a dramatic departure from the training data. We demonstrate that our NN model is capable of generalizing beyond its training data to reproduce the essential features of these much simpler scenarios. This enables us to identify strengths and weaknesses of the NN model's performance, providing us with a powerful diagnostic tool that can be used for future model optimization. These results also confirm that the NN has inferred general physical principles from its complex training data.
    
\section{Cosmic structure formation}
    \label{sec:lss}

    The evolution of cold dark matter under the influence of cosmic expansion and Newtonian gravity is typically modeled with N-body simulations. Such simulations evolve a generic, random Gaussian configuration of the initial matter field into complex structures of collapsed objects and voids that form the cosmic web at late times. Such simulations involve integrating the equations of motion of $\sim10^6$--$10^{12}$ N-body particles for thousands of time steps. Simpler models and idealized scenarios are often employed to interpret the outcomes of these numerical simulations. Here we will review both the generic nonlinear dynamics and some of these idealized scenarios in the context of Lagrangian displacement field theory (see \cite{Bernardeau:2001qr} for further details). Later we will assess the performance of our CNN model and compare with calculations done in the context of these simple scenarios.
    
    At early times, when density perturbations are small, we can describe the matter as a set of equal-mass fluid cells with time dependent positions,
    \begin{align}
        \label{eq:x}
        \mathbf{x}(a, \mathbf{q}) = \mathbf{q} + \mathbf{\Psi}(a, \mathbf{q}) \, , \\
        \mathbf{v}(a, \mathbf{q}) = \mathbf{\dot{\Psi}}(a, \mathbf{q}) \, .
    \end{align}
    Here we use the scale factor of the Universe $a$ as the time variable. The coordinates $\mathbf{q}$ are known as Lagrangian coordinates, and $\mathbf{\Psi}(a, \mathbf{q})$ is the Lagrangian displacement field. Under a Helmholtz decomposition, the displacement field is expressed as the sum of the gradient of a scalar potential $\Phi(a, \bf{q})$ and the curl of a vector potential $\mathbf{B}(a, \bf{q})$,
	\begin{align}
	    \mathbf{\Psi}(a,\mathbf{q}) = -\mathbf{\nabla} \Phi(a, \mathbf{q}) + \mathbf{\nabla} \times \mathbf{B}(a, \mathbf{q}) \, ,
	 \end{align}
	 The vector potential's divergence vanishes, $\nabla \cdot \bf{B}(a, \bf{q})=0$. The spatial derivatives are with respect to $\mathbf{q}$. Consider some early time, $a_i$, when the linearized theory accurately captures the full nonlinear dynamics. The linearized equations, known as the Ze'ldovich approximation (ZA), for each component have two solutions. The scalar potential has a growing solution $\Phi_{+}(a_i, \mathbf{q})$, and a decaying solution $\Phi_{-}(a_i, \mathbf{q})$. The vector potential has a constant solution $\mathbf{B}_{0}(\mathbf{q})$ and a decaying solution $\mathbf{B}_{-}(a_i, \mathbf{q})$. The linearized phase space then has the general form,
	 \begin{align}
	    \mathbf{x}(a_i, \mathbf{q}) = \ & \mathbf{q} + \mathbf{\nabla} \times \mathbf{B}_{0}(\mathbf{q}) -\mathbf{\nabla} \Phi_+(a_i, \mathbf{q}) -\mathbf{\nabla} \Phi_-(a_i, \mathbf{q}) \nonumber \\ & + \mathbf{\nabla} \times \mathbf{B}_{-}(a_i, \mathbf{q}) \, , \\
	    \mathbf{v}(a_i, \mathbf{q}) = & -\mathbf{\nabla} \dot{\Phi}_+(a_i, \mathbf{q}) -\mathbf{\nabla} \dot{\Phi}_-(a_i, \mathbf{q}) + \mathbf{\nabla} \times \dot{\mathbf{B}}_{-}(a_i, \mathbf{q}) \, . 	    
	 \end{align}
	 Under the ZA there is no coupling between different modes.
	 
    If $\Phi_-(a_i, \mathbf{q})$, $\mathbf{B}_0(\mathbf{q})$, and $\mathbf{B}_-(a_i, \mathbf{q})$ vanish, then the Lagrangian coordinates $\mathbf{q}$ represent the initial positions of the fluid elements in the asymptotic past. Otherwise, $\mathbf{q}$ are reference coordinates that impose some labelling scheme on the fluid elements. This description contains some redundancy, or gauge invariance. We are free to define new labels $\tilde{\mathbf{q}} = \mathbf{q} + \mathbf{\nabla} \times \mathbf{B}_{0}(\mathbf{q})$, absorbing the constant curl solution. Thus $\mathbf{B}_{0}(\mathbf{q})$ represents a gauge mode, or coordinate choice, rather than a dynamical degree of freedom, and we will drop it. This is consistent with the underlying fluid description of the system, which has four degrees of freedom: density, velocity divergence, and the two independent components of the velocity curl, or vorticity. The time derivative of the density is constrained by mass conservation to be related to the velocity divergence.

    In cosmology, the decaying solutions are neglected since they quickly become negligible. Our model's input training data is thus defined with $\Phi_-(a_i, \mathbf{q}) = \mathbf{B}_-(a_i, \mathbf{q}) = 0$. We will restrict to this class of initial conditions for now. It is convenient to consider the Fourier modes of the displacement potentials,
	 \begin{align}
	    \Phi(a, \mathbf{k}) = \int d^3 q \, \Phi(a, \mathbf{q})\, \exp(i \mathbf{k} \cdot \mathbf{q}) \, , \\
	    \mathbf{B}(a, \mathbf{k}) = \int d^3 q \, \mathbf{B}(a, \mathbf{q})\, \exp(i \mathbf{k} \cdot \mathbf{q})
	 \end{align}
	 Here we use the same symbol of the potentials and their Fourier transforms. The Fourier modes will always appear with arguments $\mathbf{k}$ rather than $\mathbf{q}$ to indicate which one is meant. Nonlinear gravitational clustering couples the modes together, so that the late time solutions to the field equations have the form,
	 \begin{align}
	    \label{eq:nlphi}
	     \Phi(a, \mathbf{k}) = \ & G_1(a,\mathbf{k}) \Phi_+(a_i, \mathbf{k}) \nonumber \\ 
	     + & \sum_{n = 2}^{\infty} \int G_n(a,\mathbf{k}_1,...,\mathbf{k}_n, \mathbf{k}) \prod_{j=1}^{n} \dee^3 k_j \Phi_+(a_i, \mathbf{k}_j) \, , \\
	     \label{eq:nlA}
	     \mathbf{B}(a, \mathbf{k}) = & \sum_{n = 3}^{\infty} \int \mathbf{G}_n(a,\mathbf{k}_1,...,\mathbf{k}_n, \mathbf{k}) \prod_{j=1}^{n} \dee^3 k_j \Phi_+(a_i, \mathbf{k}_j) \, .
	 \end{align}
	 The integration kernels $G_n$ and $\mathbf{G}_n$ can be computed as Green's functions under perturbation theory. These Green's functions also depend on cosmological parameters, but since we restrict to a single cosmology here we neglect this dependence. In the fully nonlinear regime all terms in the sums become important and the perturbative approach breaks down. Simulations are then required to model the solution. 
	 
	 Note that the nonlinearities in the above equations are convolutions, similar to the convolutional operations carried out by the CNN, so we expect the model to be able to encode the form of these Green's functions. The potential modes for N-body simulations are initialized as a Gaussian random field,
	 \begin{align}
	     \label{eq:ics}
	     \Phi_+(a_i, \mathbf{k}) = \frac{\sigma(a_i, k)}{k^2} \big(\hat{N}_{1}(\mathbf{k}) + i \hat{N}_{2}(\mathbf{k})\big) \, ,
	 \end{align}
	 where $\hat{N}_{1}$ and $\hat{N}_{2}$ are unit Gaussian random variables, and the standard deviation for the mode amplitudes is given in terms of the matter power spectrum $P_{\mathrm{mm}}(a_i, k)$ and the box volume $V$,
	 \begin{align}
	     \sigma(a_i, k) = \sqrt{\frac{P_\mathrm{mm}(a_i, k)\, V}{2}} \, .
	 \end{align}
	The Gaussian random fields used as training data contain a large sample of the mode couplings, allowing the model to thoroughly learn the forms of the nonlinear integration kernels. We demonstrate this by testing our model on initial data containing only one or two modes.
	
	 More generally, we could include decaying modes in the initial conditions, which would lead to additional terms in the evolution. For example, the scalar potential would contain mode couplings of the form,
	 \begin{align}
	    \label{eq:tker}
	    \int & G_{lmn,i_1...i_l}(a,\mathbf{k}_1,...,\mathbf{k}_{l+m+n}, \mathbf{k})
        \prod_{j=1}^{l} \dee^3 k_j B_-^{i_j}(a_i, \mathbf{k}_j) \nonumber \\ &
	    \prod_{j=l+1}^{l+m} \dee^3 k_j \Phi_-(a_i, \mathbf{k}_j) 
	     \prod_{j=l+m+1}^{l+m+n} \dee^3 k_j \Phi_+(a_i, \mathbf{k}_j) \, .
	 \end{align}
	 Here, $i_j$ are the component indices of the vector potential, and these are implicitly summed over along with the corresponding indices of the integration kernel. Similar couplings between all initial modes contribute to the vector potential as well. Since the decaying modes are excluded from our model's input data, the neural networks never have the opportunity to infer the forms of these kernels during training. We demonstrate this by testing our model on initial data containing only transversal modes.

	\subsection{\textbf{Spherical evolution}}
	\label{ssec:lss:se}

    For spherically symmetric density distributions centered around $\mathbf{q} = 0$, we can use mass conservation to derive the evolution. The initial radius Lagrangian $R$ and later Eulerian radius $r$ are related,
    \begin{align}
        R = r\, \big(1 + \bar{\delta}_{\mathrm{m}}(a, r)\big)^{\frac{1}{3}} \, ,
    \end{align}
    where the density as a function on radius is $\rho_{\mathrm{m}}(a, r) = \bar{\rho}_{\mathrm{m}}(a)\big(1 + \delta_\mathrm{m}(a, r)\big)$, and the mean overdensity within Eulerian radius $r$ is, 
	\begin{align}
	    \bar{\delta}_{\mathrm{m}}(a, r) = \frac{3}{r^3} \int_0^r \dee \tilde{r}\, \tilde{r}^2\, \delta_m(a, \tilde{r}) \, .
	\end{align}
     In this case, Newtonian gravity determines the evolution of the mean density,
	\begin{align}
	    \label{eq:se}
	    \bar{\delta}_{\mathrm{m}}''(a, r) = - & \left(2 + \frac{H'(a)}{H(a)}\right) \bar{\delta}_{\mathrm{m}}'(a, r) \nonumber \\ & + \frac{3}{2} \Omega_{\mathrm{m}}(a) \big(1 + \bar{\delta}_{\mathrm{m}}(a, r) 
	    \big) \bar{\delta}_{\mathrm{m}}(a, r) \nonumber \\ & + \frac{4}{3} \frac{(\bar{\delta}_{\mathrm{m}}'(a, r))^2}{1+ \bar{\delta}_{\mathrm{m}}(a, r)} \, .
	\end{align}
	Here primes denote derivatives with respect to $\log(a)$, $H(a) = \dee \log(a) / \dee t$ is the Hubble expansion rate of the universe, and $\om(a)$ gives the fraction of the energy density of the Universe in matter. This equation can be numerically integrated to solve for the time evolution of spherically symmetric density profiles. In the case of Einstein-de Sitter (EdS) space, where $\om = 1$, this equation has a well known analytic, parametric solution \cite{Bernardeau:2001qr}. These dynamics are fully nonperturbative, and we will demonstrate that our neural network reproduces them well when tested on initial conditions containing spherical density configurations.

    \subsection{\textbf{Isolated plane wave}}
	\label{ssec:lss:pw}
    
    Another simple scenario that admits an analytic solution is an isolated plane wave,
    \begin{align}
        \mathbf{\Psi}(a, \mathbf q) = \mathbf{A}(a) \sin(\mathbf{k} \cdot \mathbf{q} + \phi) \, .
    \end{align}
    For this configuration, the time dependence of the wave's amplitude satisfies the linearized displacement field equation of motion. Decomposing the amplitude into the longitudinal component $A_{\parallel} = \mathbf{k} \cdot \mathbf{A}$ and transverse components $\mathbf{A}_{\perp} = \mathbf{k} \times \mathbf{A}$, these satisfy
    \begin{align}
        \label{eq:zapa}
        A_{\parallel}''(a) + \left(2 + \frac{H'(a)}{H(a)}\right) A_{\parallel}'(a) = \frac{3}{2}  \Omega_{\mathrm{m}}(a)  A_{\parallel}(a) \, , \\
        \label{eq:zape}
        \mathbf{A}_{\perp}''(a) + \left(2 + \frac{H'(a)}{H(a)}\right) \mathbf{A}_{\perp}'(a) = 0. \, .
    \end{align}
    These equations can be solved analytically in $\lcdm$ in terms of hypergeometric functions. As mentioned earlier, the longitudinal component has a growing and a decaying solution, while the transverse components have a constant and a decaying solution. After dropping the constant solutions and decaying solutions, this evolution corresponds to the $G_1$ term from Eq. (\ref{eq:nlphi}). The above linearized displacement field equations have the explicit form of the ZA, so for isolated plane waves the ZA is exact. Since there is no mode coupling under the ZA, isolated modes do not couple nonlinearly with themselves. However, the discretized N-body dynamics does lead to the generation of higher modes when the initial conditions contain only a single plane wave. We will see the the neural network has learned this discretized dynamics from its training data. 
    
    \subsection{\textbf{Plane wave pairs}}
	\label{ssec:lss:wp}

    The last simple scenario that we consider is a system of two plane waves interacting. In this case there is no analytic solution except in the trivial situation where the plane waves are coplanar, in which case the modes do not couple, so they behave as isolated plane wave solutions. Due to the nonlinearity of the equation of motion, two modes that are not coplanar couple together and dynamically generate additional modes. If the displacement mode amplitudes are initially small, these nonlinearities can be computed under Lagrangian perturbation theory (LPT). As the system of modes continues to grow, eventually perturbation theory will fail to accurately capture the full nonlinear dynamics, and we must rely on N-body simulations to model the evolution. 
    
    For two initial plane waves,
     \begin{align}
        \mathbf{\Psi}(a_i, \mathbf q) = \mathbf{A}_1(a_i) \sin(\mathbf{k}_1 \cdot \mathbf{q} + \phi_1) + \mathbf{A}_2(a_i) \sin(\mathbf{k}_2 \cdot \mathbf{q} + \phi_2) \, .
    \end{align}
    the lowest order modes generated by their coupling have wave vectors $\mathbf{k}_{\pm} = \mathbf{k}_1 \pm \mathbf{k}_2$. For simplicity, we will set both initial phases to zero and neglect any initially decaying linear solutions. Then the two waves are purely longitudinal, and the waves they generate under second order LPT (2LPT) are,
    \begin{align}
        \mathbf{\Psi}_{\pm}(a, \mathbf q) = \frac{A_{\pm\parallel}(a)}{2} \left(1 - \frac{(\mathbf{k}_1\cdot \mathbf{k}_2)^2}{k_1^2 k_2^2} \right) \frac{\mathbf k_{\pm}}{k_{\pm}^2} \sin(\mathbf{k}_\pm \cdot \mathbf{q}) \, .
    \end{align}
    This lowest order mode coupling is equivalent to the $G_2$ kernel from Eq.  (\ref{eq:nlphi}). The 2LPT growth equation for the amplitudes of these generated modes is
    \begin{align}
        \label{eq:A2}
        A_{\pm\parallel}'' + \left(2 + \frac{H'(a)}{H(a)}\right) A_{\pm\parallel}' = \frac{3}{2}  \Omega_{\mathrm{m}}(a) \Big(A_{\pm\parallel} + A_{1\parallel} A_{2\parallel}\Big) \, .
    \end{align}
    Given the linear solution for the evolution of $A_{1\parallel}$(a) and $A_{2\parallel}$(a), this equation can be numerically integrated to determine the amplitudes of the 2LPT modes in the quasilinear regime.
    
    The nonlinear displacement field can be further expanded to higher order in LPT, where modes with wave number $\mathbf{k}_{mn} = m \mathbf{k}_1 + n \mathbf{k}_2$ are generated for integer values of $m$ and $n$. At higher order both longitudinal and transversal modes are produced with amplitudes that satisfy equations similar in form to Eq. (\ref{eq:A2}), sourced by products of lower order amplitudes. The Lagrangian description of the system breaks down at shell crossing, when the Jacobian becomes singular, so the mapping between Eulerian and Lagrangian coordinates is no longer well defined. We will evaluate our model with input modes that have not yet undergone shell crossing, so the perturbative description is still valid and the mode coupling is well understood.
    
	\begin{figure*}
		\centering
		\graphic{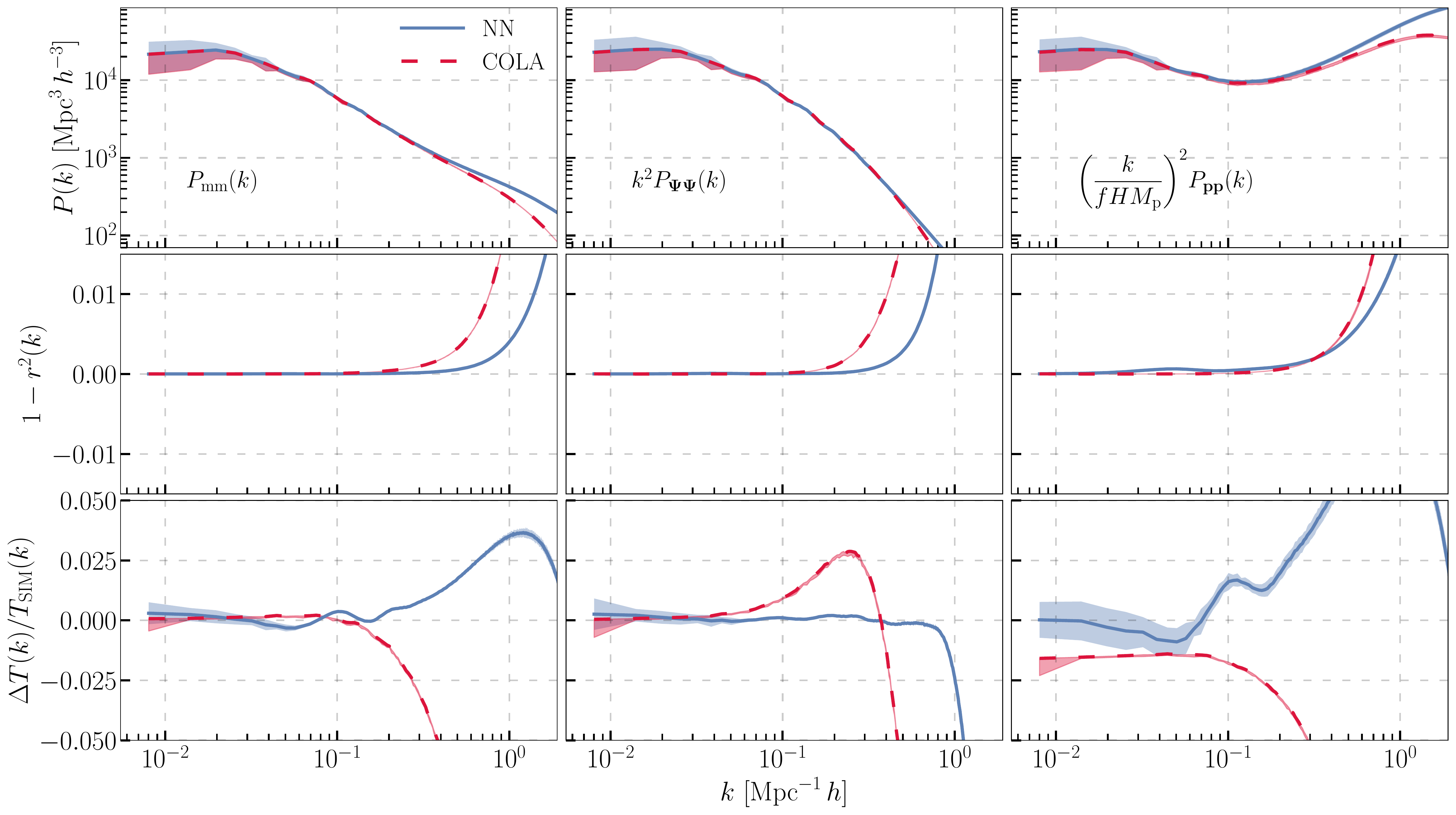}{0.95}
		\caption{The top row shows the matter density power spectra (left) from the CNN displacement model (blue, solid) and COLA (red, dashed), the Lagrangian displacement field (middle), and the momentum field (right). The momentum field depends on both particle displacements and velocities. The displacements and momenta have been rescaled to coincide with the matter power spectrum on large scales. The middle row shows the stochasticities with respect to the N-body simulations, and the bottom row shows the relative error in the transfer functions.}
		\label{fig:Pe}
	\end{figure*}

	\section{Results}
	\label{sec:resu}

	\subsection{\textbf{Power spectra}}
	\label{ssec:resu:pk}

	One of the most important cosmological observables is the matter density power spectrum. For a density field $\delta_{\mathrm{m}}(a, \mathbf{x})$, the Fourier modes of this field are given by
	\begin{align}
	    \delta_{\mathrm{m}}(a, \mathbf{k}) = \int \dee^3 x \,  \delta_{\mathrm{m}}(a, \mathbf{x}) \exp(i\mathbf{k}\cdot\mathbf{x}) \, .
	\end{align}
	Then the density power spectrum, $P_{\mathrm{mm}}(a, k)$ is defined in terms of the equal-time, two-point correlation function of these Fourier modes,
	\begin{align}
	    \langle \delta_{\mathrm{m}}(a, \mathbf{k}) \delta_{\mathrm{m}}(a, \mathbf{k}') \rangle = (2\pi)^3 \delta_{\mathrm{D}}^{(3)}(\mathbf{k} + \mathbf{k}') P_{\mathrm{mm}}(a, k) \, .
	\end{align}
	On large scales (small $k$), where the evolution is well described by the ZA, the shape of the power spectrum encodes information about the physics of the early universe and the statistics of the primordial density perturbations. On small scales, nonlinear evolution, along with baryonic effects, make it difficult to predict the shape of the power spectrum and extract cosmological information. For this reason it is important to have robust and accurate modeling of the full nonlinear evolution of the matter fluctuations on small scales. Here, we use the small-scale power spectrum as a basis of comparison against both the full N-body evolution, and the COLA approximation.
	
	In addition to the matter power spectrum, we also consider the Lagrangian displacement power spectrum.  This is more directly related to the output of the NN, although less relevant for deriving constraints from observational surveys. The Fourier modes of the displacement field are given by
	\begin{align}
	    \mathbf{\Psi}(a, \mathbf{k}) = \int \dee^3 q\, \mathbf{\Psi}(a, \mathbf{q})  \exp(i\mathbf{k}\cdot\mathbf{q}) \, ,
	\end{align}
	and their power spectrum is defined
	\begin{align}
		\langle \mathbf{\Psi}(a, \mathbf{k}) \cdot \mathbf{\Psi}(a, \mathbf{k}') \rangle = (2\pi)^3 \delta^{(3)}_{\mathrm{D}}(\mathbf{k} + \mathbf{k}') P_{\mathbf{\Psi\Psi}}(a, k) \, .
	\end{align}

    To test the accuracy of the velocity model we construct the Eulerian momentum field $\mathbf{p}(a, \mathbf{x})$. From the Fourier modes of this field, the momentum power spectrum is given by
  	\begin{align}
		\langle \mathbf{p}(a, \mathbf{k}) \cdot \mathbf{p}(a, \mathbf{k}') \rangle = (2\pi)^3 \delta^{(3)}_{\mathrm{D}}(\mathbf{k} + \mathbf{k}') P_{\mathbf{pp}}(a, k) \, .
	\end{align} 

    In practice, since the N-body particles represent a discrete point distribution of mass and momentum, the Eulerian density and momentum fields must be constructed by smoothing the particles onto a mesh. We used the cloud-in-cell (CIC) mesh assignment scheme for this purpose on a mesh with $1024^3$ grid sites. The mesh is then Fourier transformed using FFTW3 \cite{Frigo:2005zln}, and the resulting density modes are binned according to wave number in bins the width of the fundamental mode in the simulation box. The Eulerian power spectra are computed as the average square amplitude in each wave number bin. The Lagrangian displacement power spectrum is Fourier transformed directly, since the displacement field is already defined with respect to a regular grid. In this case, the mesh size is $512^3$, which is the same as the number of simulation particles. The displacement power spectrum is then computed using the same wave number binning as the Eulerian power spectra.
	
	For each quantity $\delta_{\mathrm{m}}$, $\mathbf{\Psi}$, and $\mathbf{p}$, we compute their modes from the CNN model output, the COLA simulations, and the N-body simulations. We characterize the errors in our CNN predictions by computing the stochasticity,
	\begin{align}
		1 - r^2(k) \equiv 1 - \frac{P_{\mathrm{NN} \times \mathrm{SIM}}(k)}{\sqrt{P_{\mathrm{NN}}(k) P_{\mathrm{SIM}}(k)}} \, ,
	\end{align}
	where $P_{\mathrm{NN} \times \mathrm{SIM}}$ is the cross power spectrum between the model prediction and the simulation modes, $P_{\mathrm{NN}}$ is the auto power spectrum from the CNN prediction, and $P_{\mathrm{SIM}}$ is the auto power spectrum from the simulation. The stochasticity quantifies the excess correlation in the predicted data that does not match the correlations in the simulation data. We also compute the error in the transfer functions of the CNN output relative to the simulations. For the CNN, the relative error in the transfer functions are,
	\begin{align}
		\frac{\Delta T(k)}{T_{\mathrm{SIM}}(k)}  = \sqrt{\frac{P_{\mathrm{NN}}(k)}{P_{\mathrm{SIM}}(k)}} - 1 \, .
	\end{align}	
    We similarly compute the stochasticity and transfer function errors from the COLA simulations relative to the N-body results.

	In Fig. \ref{fig:Pe}, we plot the power spectra along with their stochasticities and transfer function errors. We have rescaled the displacement and momentum power spectra to coincide with the density power spectrum on large scales. We computed these results for ten realizations of the linear density field that were not included in the CNN training or validation data sets. The shaded regions in Fig. \ref{fig:Pe} show the $2\sigma$ intervals estimated by bootstrap averaging over these ten realizations.
	
	Our CNN model achieves an accuracy of a few percent at deeply nonlinear scales of $k\sim 1\ \impc\, h$. Additionally, the CNN model significantly outperforms COLA on these nonlinear scales. In terms of computational speed, our model is comparable to COLA, although for optimal performance it must be run on GPU. In terms of accuracy, our models are comparable to the full N-body evolution on $\mpc/h$ scales.
	
	The momentum CNN errors are larger than the displacement errors. This is because the distribution of the particles onto the Eulerian mesh depends on the particle positions. We use the output of the displacement CNN for this purpose, so the displacement errors propagate to the momentum Fiels. The CNN transfer function errors have small oscillations on scales where $k \sim 0.1\,\impc\, h$, which can most easily be seen in the momentum power spectrum. These correspond to slight errors in reproducing the baryonic acoustic oscillations (BAO) of the power spectrum.
	
	\subsection{\textbf{Spherical evolution tests}}
	\label{ssec:resu:se}

	\begin{figure*}
    	\centering
    	\graphic{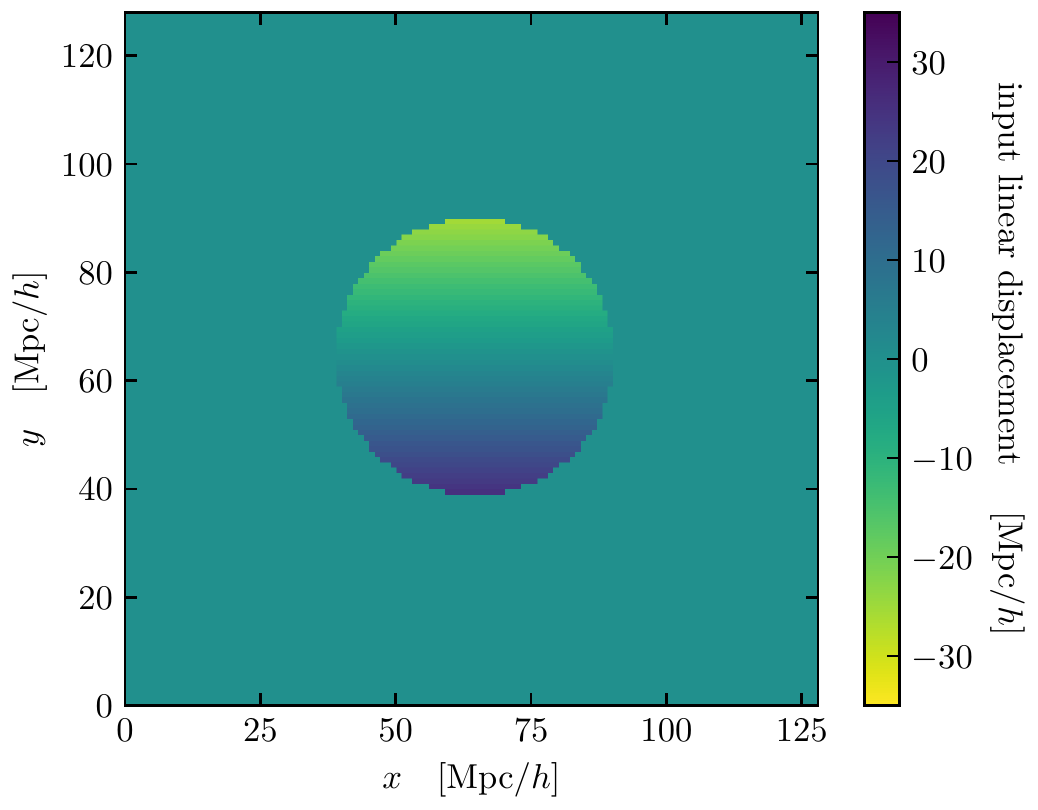}{0.45}
    	\qquad
    	\graphic{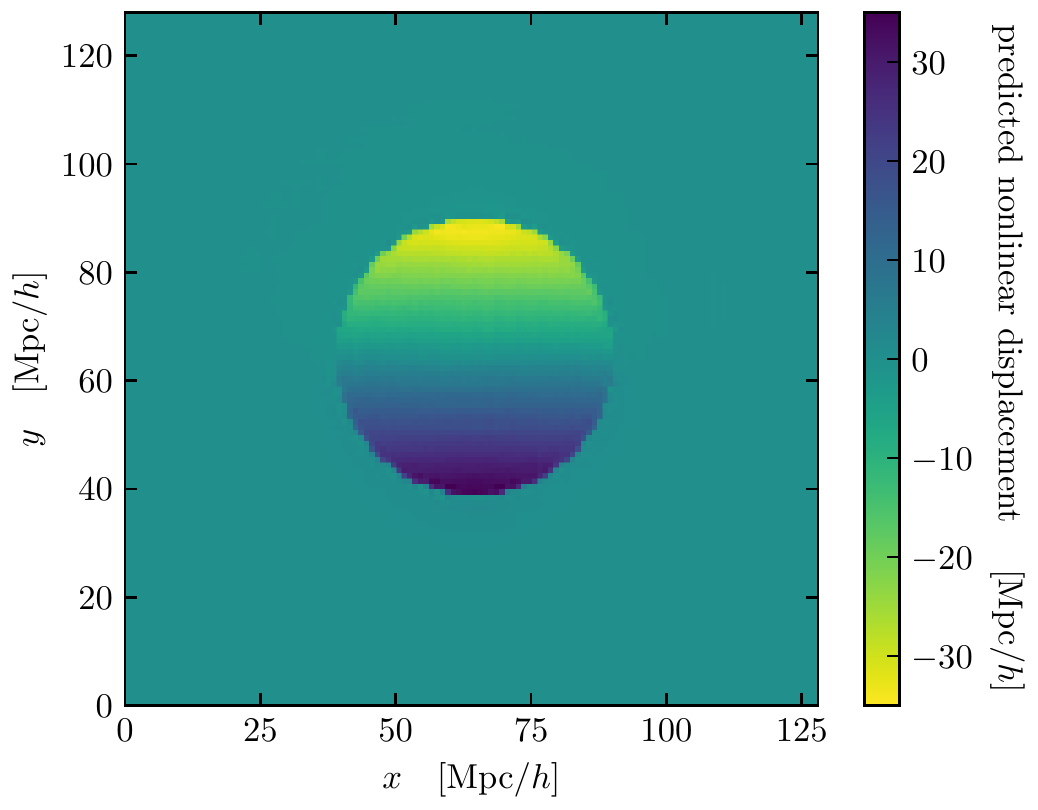}{0.45}
    	\caption{Spherical evolution test displacement fields. Our neural network models take linear displacement fields as input and predict nonlinear outputs to emulate N-body simulations. We consider test cases with spherical symmetry that admit exact solutions. The left panel shows the linearly varying $y$ displacement of particles in a 2D slice through the center of a $50\,\mpc\,h^{-1}$ sphere in the Lagrangian space. The inner particles are set up in a linear overdensity of 1.55 so they collapse into a smaller region at redshift $z=0$ without shell crossing. The right panel shows the nonlinear network prediction of the same field. The nonlinear displacements also vary linearly, and are much larger in magnitude to create a nonlinear density of $\approx20$.}
    	\label{fig:sphere}
    \end{figure*}
    
	\begin{figure}[tb]
		\centering
		\graphic{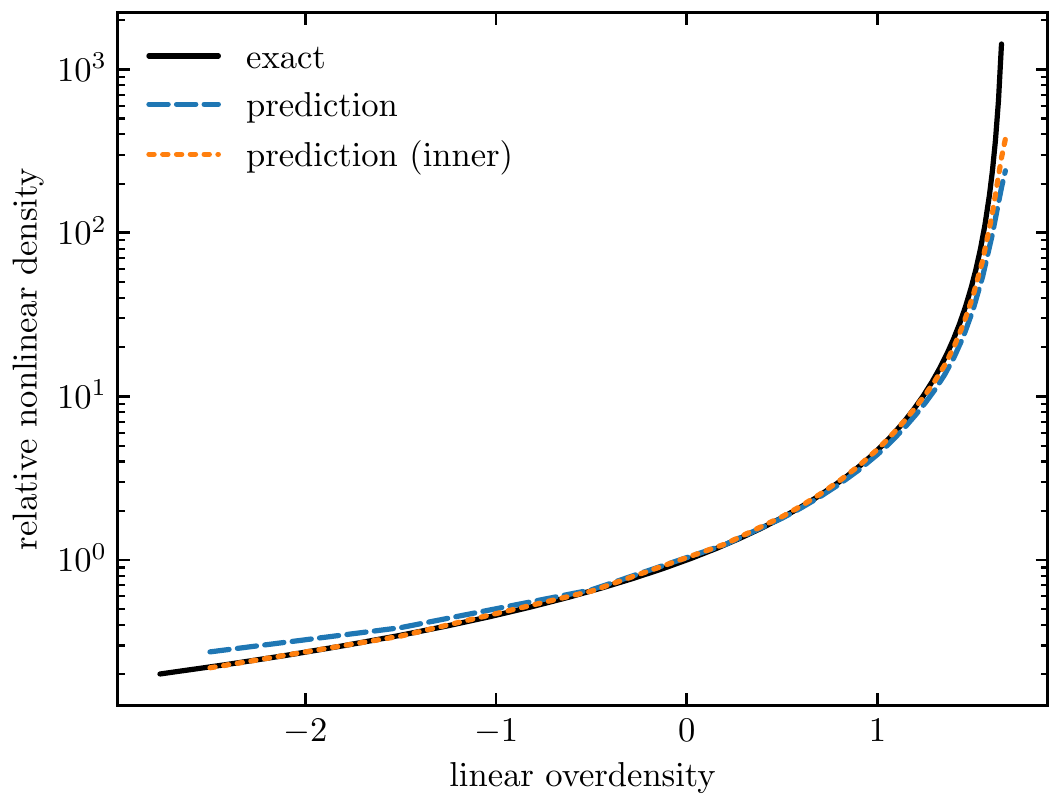}{0.95}
		\caption{Spherical evolution density relation. Underdense spherical top hats expand and overdense ones collapse. Depending on their linear overdensities, the top hats evolve into different nonlinear densities at redshift 0, which can be solved exactly before shell crossing. We compare the neural network predictions to the exact solutions and find good agreement, which is further improved if we avoid the fuzzy edges by only using the inner part (within 0.8 times the radius) of the sphere to estimate the nonlinear relative density. The fuzzy edges can be seen in the right panel of Fig.~\ref{fig:sphere}.}
			\label{fig:sphere_dens}
	\end{figure}

	Analytically solvable models offer rich physical insights without handling the full complexity of a problem. They typically impose symmetries on the system configuration, as in the spherical evolution model in cosmology. In a matter dominated universe (EdS), the collapse of overdense (as compared to the mean density of the universe) spheres have analytic solutions before shell crossing, as does the expansion of spherical underdense regions. Having trained our neural network models with complex systems from simulations, we can use the exact solutions as test cases to inspect their performance and understand how they work.

	For this purpose, we set up spherically symmetric ZA inputs of radius $50\,\mpc\, h^{-1}$ in $250\,\mpc\, h^{-1}$ boxes. The boxes are divided on $128^3$ grids so that their cell size is the same as in the N-body simulations. Note that by design our models can apply to different volumes of the same grid resolution. We choose a list of spherical top hats with uniform density contrasts $\delta_\mathrm{sphere} \in \{$-2.5, -1.5, -0.5, 0.2, 0.5, 0.6, 0.7, 0.8, 0.9, 1.0, 1.05, 1.1, 1.15, 1.2, 1.25, 1.3, 1.35, 1.4, 1.45, 1.5, 1.55, 1.6, 1.65, 1.67$\}$, in which the underdense ($\delta_\mathrm{sphere} < 0$) top hats expand and the overdense ($\delta_\mathrm{sphere} > 0$) ones collapse. The values span almost the full range of scenarios between expansion shell crossings at the boundary and collapsing shell crossings at the center \cite{ShethVanDeWeygaert2004}.

	For each test case we apply the displacement CNN model to the ZA inputs. Fig.~\ref{fig:sphere} shows the $y$ displacements of particles in a 2D slice through the center of the sphere with $\delta_\mathrm{sphere}=1.55$ as an example. The linear input displacements, shown on the left of Fig.~\ref{fig:sphere}, vary linearly in the $y$ direction, showing that the particles move towards the center. On the right, the CNN predicted displacements are significantly larger, resulting in a nonlinear overdensity of $\approx20$. Displacements with other values of $\delta_\mathrm{sphere}$ are qualitatively similar.

	Quantitatively, we compare the nonlinear densities inside the spherical top hats to the numerically integrated solution of Eq. (\ref{eq:se}) for the same $\lcdm$ cosmology as of the simulations. We measure the inner densities of the model outputs using the initial linear positions and final nonlinear positions of the particles. Let $\bar{r}_{\mathrm{lin}}$ be the average radial distance of the particles from the center of the grid in the linear input, and let $\bar{r}_\mathrm{nl}$ be the average radius of their final positions. The relative nonlinear density relative to the background density is given by $(\bar{r}_\mathrm{lin} / \bar{r}_\mathrm{nl})^3$.
	
	The nonlinear density is a monotonic function of the linear one, and should stay uniform if the linear input is uniform. Fig.~\ref{fig:sphere_dens} shows that the CNN model accurately predicts this relation for linear overdensities between $-1$ and $1$, while having a small deviation beyond that range. This is due to the fact that the network is not fully resolving the sharp edges of the spherical top hats, as can be seen in Fig.~\ref{fig:sphere}. This causes some leakage to outer regions in the density estimation. To verify this, we also measure the nonlinear density using only particles within 0.8 of the sphere radius. The inner radius results match the exact solutions more accurately, as shown in Fig.~\ref{fig:sphere_dens}. Note that the edge of the top hat is both where the distribution of matter has the most rapid spatial variation and where the neural network errors dominate. We will return to this observation in the next subsection.

	\begin{figure*}
		\centering
		\graphic{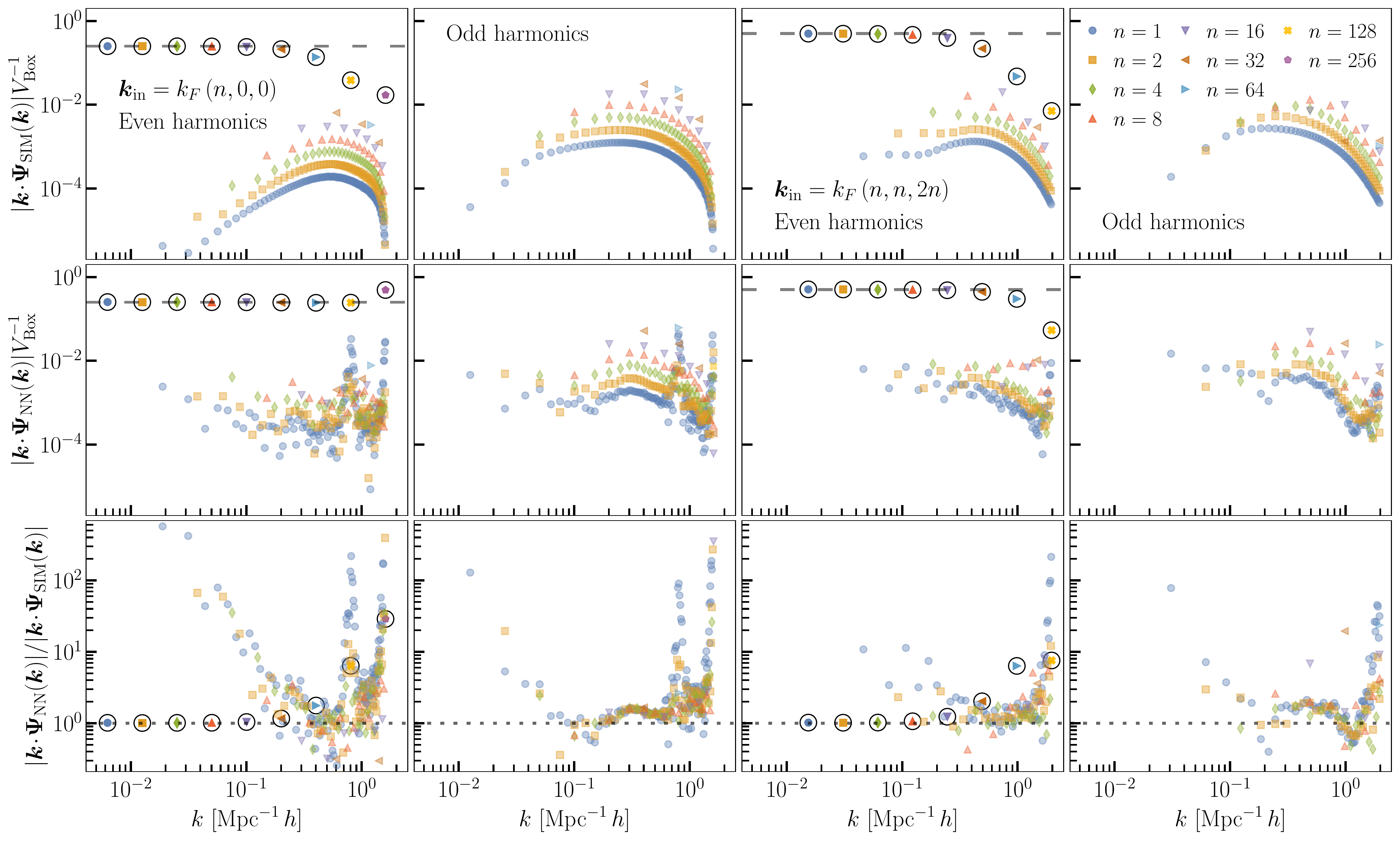}{1.}
		\caption{Displacement divergence amplitudes from isolated longitudinal modes. Top panels show the N-body simulation results, middle panels show the neural network predictions, and bottom panels show the ratio of neural network to N-body mode amplitudes. The left two columns are for longitudinal modes with displacements only along the $x$ direction. The right two columns are for modes with equal displacement amplitudes along the $x$ and $y$ directions and double these displacements in the $z$ direction. The dashed lines in the top two rows show the expected amplitudes from theory. The circled data points correspond to the original input modes in the initial conditions. The first and third columns show the input mode and even harmonic modes, while the second and fourth columns show the odd harmonics. The higher harmonics are absent in the continuum limit and arise from the discreteness of the N-body systems, which the neural network learns from the N-body training data.}
		\label{fig:sl}
	\end{figure*}

	\subsection{\textbf{Isolated longitudinal modes}}
	\label{ssec:resu:im}

    Another instructive example of a solvable system is an isolated plane wave. As described in subsection \ref{ssec:lss:pw}, for purely longitudinal ($\mathbf{\nabla} \times \mathbf{\Psi} = 0$) and purely transversal ($\mathbf{\nabla} \cdot \mathbf{\Psi} = 0$) waves of particle displacement, the ZA (Eqs. (\ref{eq:zapa}--\ref{eq:zape})) is the exact solution for their time evolution.

    In our simulation boxes, wave vectors have the form $\mathbf{k} = k_\mathrm{F}(n_x, n_y, n_z)$, where $k_\mathrm{F} = 2\pi/L_{\mathrm{Box}}$ is the fundamental wave number of the box and the $n_i$ are integers. We ran N-body simulations with ZA initial conditions containing single isolated longitudinal displacement modes with initial wave vectors $\mathbf{k}=k_{\mathrm{F}}(n, 0, 0)$ where $n \in \{1,\, 2,\, 4,\, 8,\, 16,\, 32,\, 64,\, 12,\, 256\}$. The first mode is the fundamental mode, and the last is the Nyquist mode for the 3D cubic mesh. We also ran a set of simulations with initial longitudinal modes $\mathbf{k} = k_{\mathrm{F}}(n, n, 2n)$. For this set of diagonal waves, we omit the $n=256$ case, since it is beyond the Nyquist mode in the last dimension. Each mode was given a linear amplitude corresponding to $A_\parallel = 1/2$, which ensures that shell crossing does not occur.

	The amplitudes of displacement divergence modes from the N-body simulation and the CNN are plotted in Fig. \ref{fig:sl}. The circled data points correspond to the input, ZA modes. The panels in the first and third columns show the input modes along with even order harmonics, and panels in the second and fourth columns show the odd order harmonics. These higher harmonics are present in both the N-body simulation and in the CNN output due the discreteness of the N-body evolution, which the CNN learns.

    In the let column of Fig. \ref{fig:sl}, the final amplitude of the input modes are predicted to better that 1\% accuracy for $n<16$ and to within 3\% accuracy including $n=16$, corresponding to $k=5.0\times10^{-2}\impc\, h$). This demonstrates that the CNN preserves the ZA evolution on the largest scales, as it should. The CNN overpredicts the amplitudes of smaller scale modes, approaching the deeply nonlinear regime.  In the continuum limit, the growth of these displacement modes should be completely independent of scale. However, we see that the N-body simulations exhibit a decreasing amplitude on small scales where the waves are poorly sampled by a small number of discrete particles per wavelength. In this sense, the neural network predictions are actually closer to the continuum evolution. Although, for the $x$-directed wave with the smallest wavelength, the CNN predicts an enhanced amplitude. Such a large error indicates that this mode is past the limit of the CNN's learned resolution. For the diagonal waves (the two rightmost columns in Fig. \ref{fig:sl}), the CNN amplitudes decrease with increasing $k$, better reproducing the N-body results.
	
	The CNN reproduces the higher harmonics best on scales between $0.2\,\impc\, h < k < 0.8~\impc\, h$. Outside of this range of scales, the CNN overpredicts the amplitudes. The odd harmonics have larger amplitudes than their nearest neighbor even harmonics. However, the neural network predicts them to have similar amplitudes, and so the CNN generally reproduces the N-body odd harmonics better than the even harmonics. To understand the pattern of CNN errors for higher harmonics, we must consider the waves in configuration space. As a function of initial Lagrangian position, the waves have nodes at every half wavelength where the displacement field vanishes. The distribution of matter has the highest spatial variation at these nodes. Just as in the previous subsection, the CNN errors are largest in regions of rapid density variation, so these nodes are where the CNN errors dominate. It is the even harmonics that have nodes that overlap with the input wave nodes, so these have worse errors than the odd harmonics. 

	We also ran a set of simulations in which the initial mode was fixed as $\mathbf{k}_{\mathrm{lin}} = k_F(16, 0, 0)$ and the amplitude of the mode was varied from $A_\parallel = 0.1$ to $A_\parallel = 8$ (see Eq. (\ref{eq:zapa})). Note that $A_\parallel = 1$ is the minimum amplitude for which an isolated mode has undergone shell crossing at redshift zero, since the determinant of the Jacobian between the Eularian and Lagrangian coordinates vanishes wherever $\mathbf{q}\cdot\mathbf{k} = \pi/2$ for these modes. 
	
\begin{figure}
		\centering
		\graphic{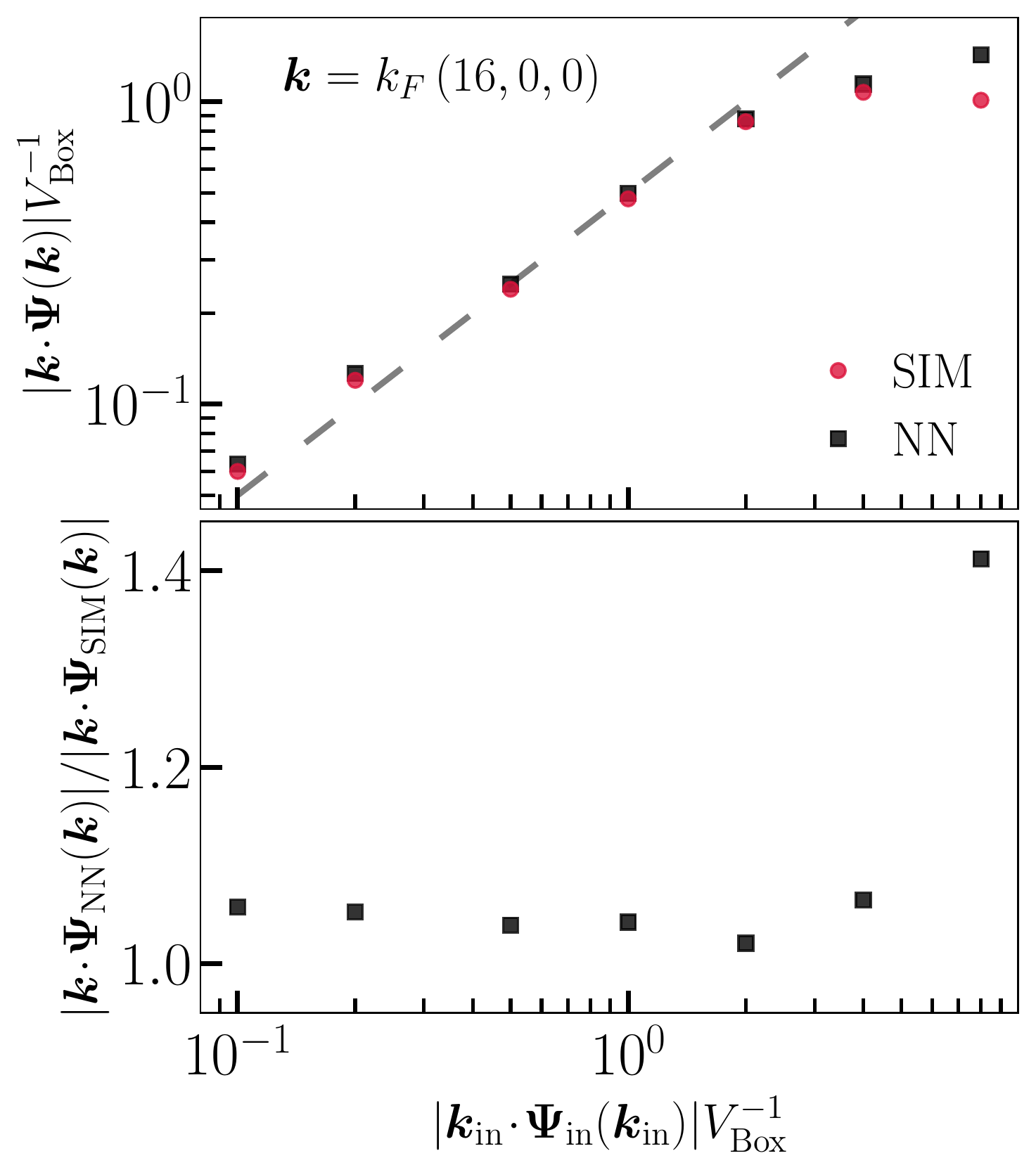}{0.7}
		\caption{Displacement divergence mode amplitudes from N-body simulations and neural network predictions for a fixed wave vector and varied initial amplitude. The dashed line shows the expected amplitude from linear theory.}
		\label{fig:sla}
	\end{figure}

	The results for the final amplitudes are shown in Fig. \ref{fig:sla}. The dashed line is the expected final amplitude from theory. Both the CNN and the N-body evolution have enhanced final amplitudes when the initial amplitudes are $A_\parallel=0.1$ and $A_\parallel=0.2$, which are the smallest amplitudes we considered. The neural network is about 6\% higher than the N-body evolution for these cases. For input amplitudes from $A_\parallel=0.5$ to $A_\parallel=2.0$, both the CNN and N-body results agree well with theory, meaning the neural network can accurately predict the amplitudes of isolated modes even after shell crossing. This indicates that the CNN has learned nonperturbative aspects of the dynamics from the training data, which is well past shell crossing. For higher initial amplitudes, the N-body results are lower and diverging away from the linear growth calculation and the CNN overpredicts these N-body amplitudes.
	
    Input data containing a single plane wave are a significant departure from the training data that the neural network has encountered. Errors in the CNN predicted amplitudes are thus expected for these extremely simple configurations, especially on small scales where nonlinear mode coupling is nonperturbative so the behavior of isolated modes may not be easily inferred in this regime. However, the CNN predicts a set of mode amplitudes that matches the full N-body evolution in most cases, even after shell crossing. This demonstrates that the CNN has inferred some general physical principles from its training data, enabling it to accurately generalize to the evolution of isolated plane waves across a wide range of scales and amplitudes.
	
	\begin{figure}
		\centering
		\graphic{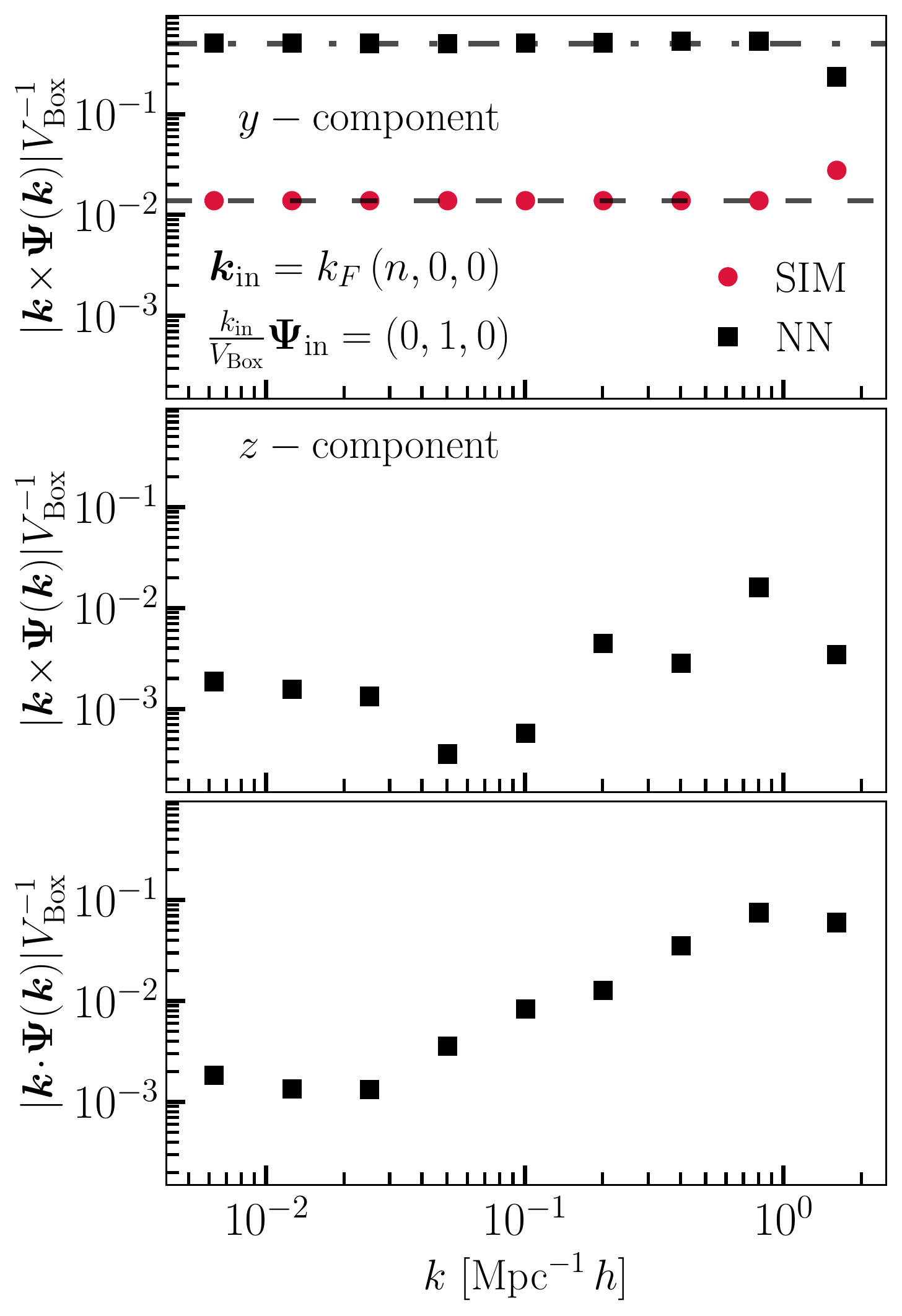}{0.7}
		\caption{Mode amplitudes from N-body simulations and neural network predictions with initial data corresponding to isolated transversal modes. The top panel shows the displacement curl component that is consistent with the initial data. The dashed line is the expected amplitudes from linear theory, and the dash-dotted line is the expected amplitude for a longitudinal mode. The second panel shows spurious curl modes predicted by the neural network that are perpendicular to the initial wave vector but parallel to the initial displacements. The bottom panel shows spurious longitudinal modes predicted by the neural network.}
		\label{fig:st}
	\end{figure}

	\begin{figure*}
		\centering
		\graphic{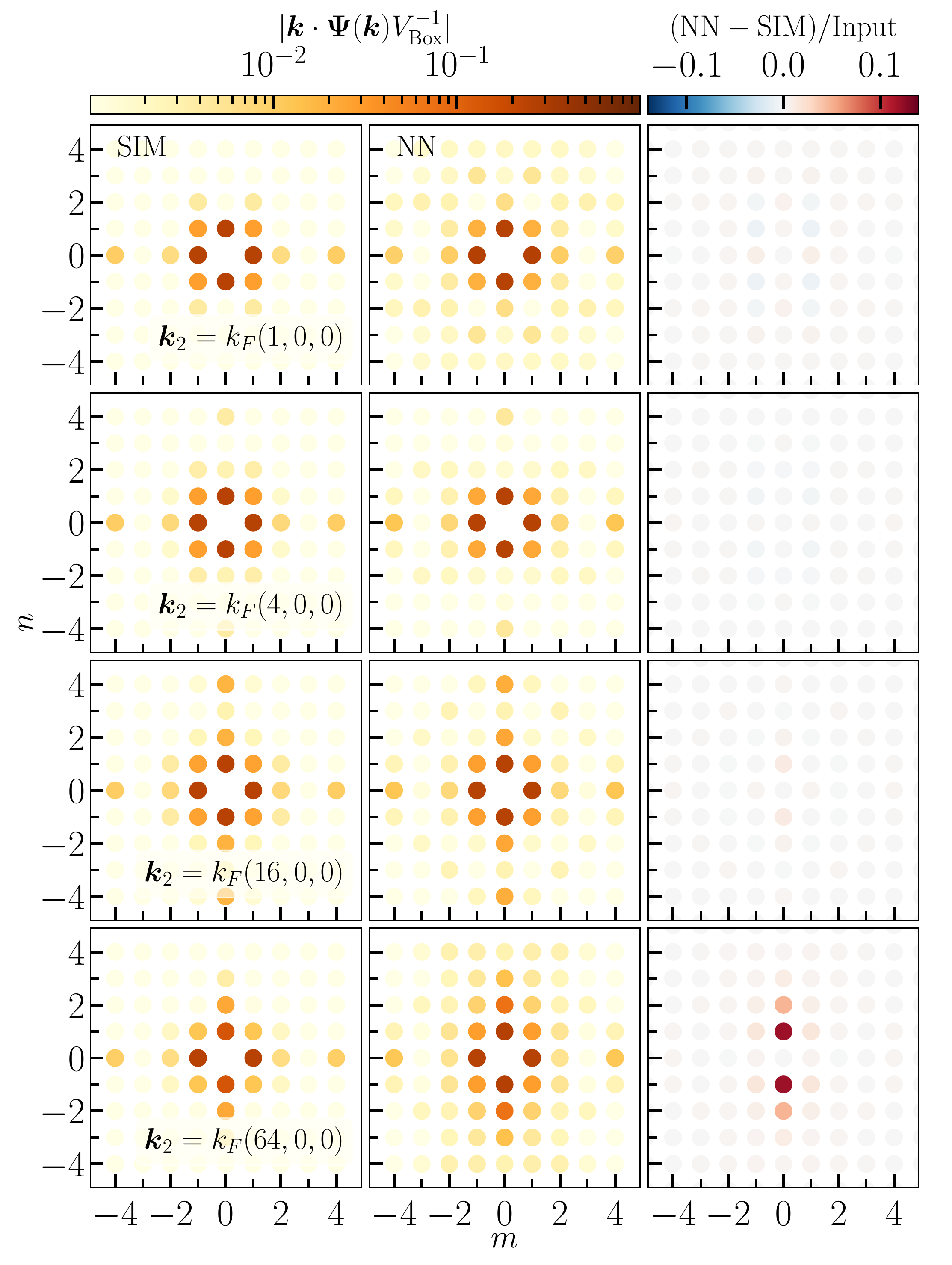}{0.495}
		\graphic{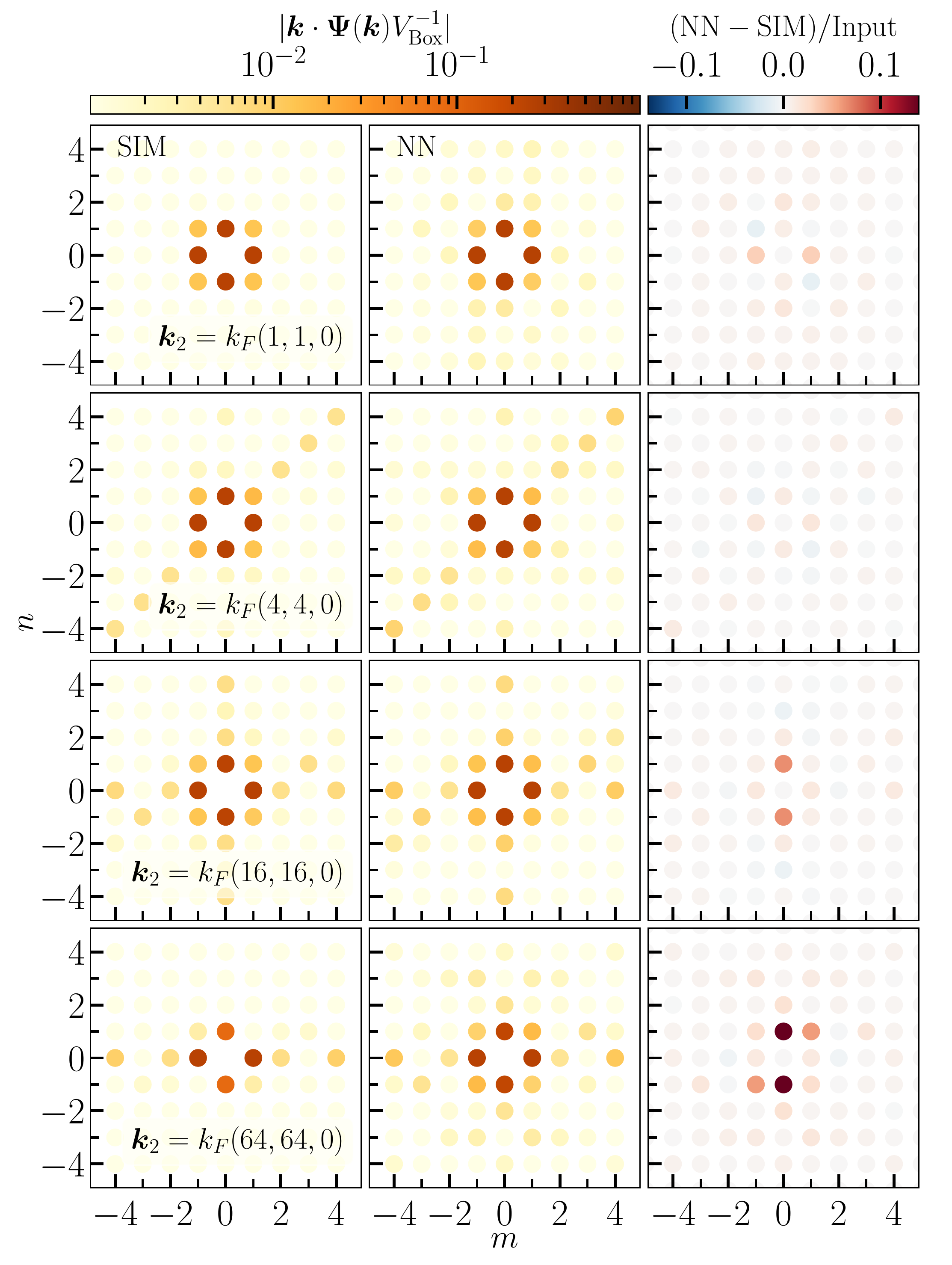}{0.495}
		\caption{Amplitudes of displacement divergence for initial data with two longitudinal waves. The left set of plots show cases where $\mathbf{k}_2$ is perpendicular to $\mathbf{k}_1$, and the right set shows cases where $\mathbf{k}_2$ has both perpendicular and parallel components to $\mathbf{k}_1$. Each set of plots shows the amplitudes from the simulations on the left, the neural network in the middle, and their difference relative to the input amplitude on the right. Each color dot corresponds to the amplitude of a mode with wave vector $\mathbf{k}_{mn} = m \mathbf{k}_1 + n \mathbf{k}_2$, where values of $m$ and $n$ are along the horizontal and vertical axes respectively.}
		\label{fig:mn_pairs}
	\end{figure*}

	\subsection{\textbf{Isolated transversal modes}}

	The two previous examples focused on simple physical systems that can be solved exactly, and for which the CNN should be able to generalize its training and make reasonable predictions. This is because these systems evolve according to the same underlying physics that the CNN, to some degree, is inferring as it trains. An example of a simple physical system that we expect the neural network to perform poorly on is a transversal displacement mode, which is totally absent from the training data.
	
	The transversal ZA, given by Eq. (\ref{eq:zape}), is the exact solution for isolated transversal displacement modes. In this case, time dependent amplitudes evolve according to a constant and a decaying solution, but no growing solution. We include both a constant and a decaying component in the transversal waves we consider, in order to give them analogous initial conditions to the longitudinal modes. Specifically, our choice of initial condition for these modes is such that the displacement and velocity have the same coefficients at the initial time as the longitudinal modes discussed earlier, with initial velocities in the same direction as their displacements. This is achieved by giving the decaying component a coefficient with the opposite sign to the time independent component. The mode decelerates as it grows in amplitude, asymptotically approaching the amplitude of the purely the time independent part. However, this is really just a physical mode decaying away and leaving behind a residual, unphysical gauge mode.
	
	Typically N-body simulations omit transversal modes, since they represent only decaying or pure gauge solutions. Thus, the CNN has never seen a transversal mode in its training data, and we do not expect it to accurately predict the evolution. We test this by running several N-body simulations for isolated waves with initial displacement amplitudes $\mathbf{A}_{\perp} = (0, 0, 1)$ and wave vectors $\mathbf{k} = k_F (n, 0, 0)$, and compare these with CNN predictions.

	The results are plotted in Fig. \ref{fig:st}. The top panel shows amplitudes of the displacement field curl modes along the direction corresponding to the curl of the input mode. The N-body simulation produces mildly growing modes that have already converged to a nearly time independent behavior. The dashed line in the top panel of Fig. \ref{fig:st} shows the expected final amplitudes from linear theory. The neural network predicts that this mode will grow exactly as a longitudinal mode (the dash-dotted line), since this is the only behavior it has encountered.
	
	In the middle panel of the same figure we show modes that are perpendicular to the expected curl of the displacement field. These are perpendicular to the initial wave vector but parallel to the initial displacement vectors. On the bottom panel we show the divergence modes. These modes vanish in the N-body results, as they should, but the CNN predicts nonvanishing amplitudes for them. Interestingly, we find that the CNN correctly predicts vanishing amplitudes for the curl component perpendicular to the initial displacements but parallel to the initial wave vector. Clearly, the CNN is unable to predict the correct evolution for transversal modes because it has never been exposed to them. In general, the model has not inferred any of the forms of integration kernels, such as those in Eq. (\ref{eq:tker}), containing initially decaying modes.

	\begin{figure*}
		\centering
		\graphic{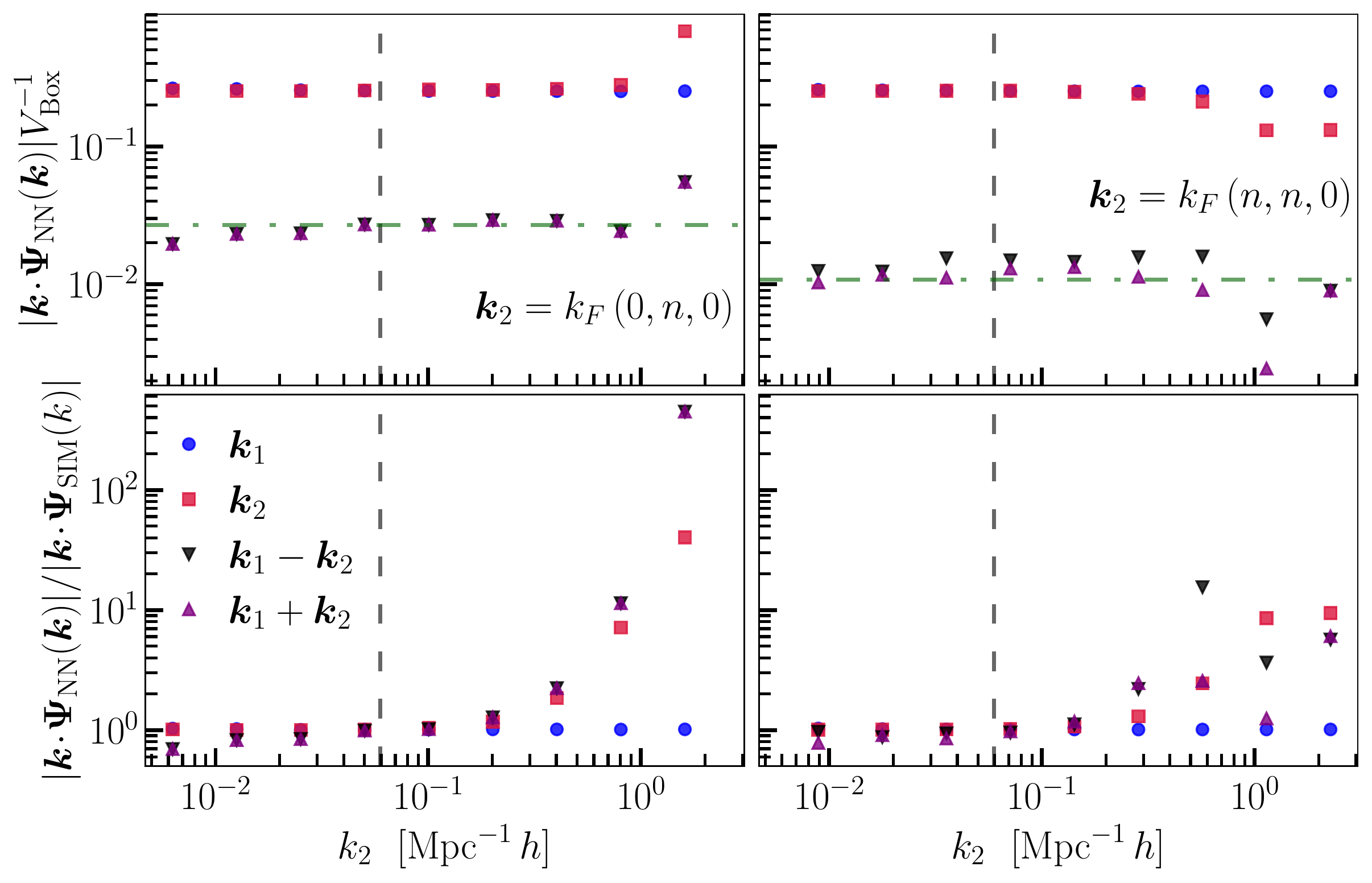}{0.6}
		\caption{Displacement divergence amplitudes for input containing two longitudinal modes. The blue circles correspond to input modes with the wave vector $\mathbf{k}_1$, which is held fixed for all the initial conditions. The red squares correspond to the input modes with wave vectors $\mathbf{k}_2$. The black and purple triangles show the amplitudes of modes that are generated in 2LPT. The green horizontal dash-dotted line shows the predicted 2LPT amplitude for these modes. The vertical dashed line shows the scale approximately corresponding to the CNNs field of view. The top row shows the CNN predictions, and the bottom row shows their ratio to the N-body simulation results.}
		\label{fig:pair_2lpt}
	\end{figure*}

	\subsection{\textbf{Coupled mode pairs}}
	
    As a final example, we consider a system initially containing two noncoplanar, longitudinal displacement modes. As discussed in subsection \ref{ssec:lss:wp}, the evolution of this system cannot be solved analytically, but can be accurately described by LPT as long as the mode amplitudes remain sufficiently small.

	We ran sets of N-body simulations for systems containing only two longitudinal waves in their initial conditions. For both sets, one mode was fixed as $\mathbf{k}_1 = k_F (8, 0, 0)$, while the second mode differed between simulations. One set had $\mathbf{k}_2$ perpendicular to $\mathbf{k}_1$, with wave vectors $\mathbf{k}_2 = k_F (0, n, 0)$ for n ranging from 1 to 256 in powers of 2. The other set had wave vectors $\mathbf{k}_2 = k_F (n, n, 0)$, with both a parallel and perpendicular component to the first wave vector.

	In Fig. \ref{fig:mn_pairs}, we show the amplitudes of the displacement divergence modes with wave vectors given by $\mathbf{k}_{mn} = m \mathbf{k}_1 + n \mathbf{k}_2$ up to fifth order in $m$ and $n$. Qualitatively we see that the largest amplitude modes of the neural network match the largest amplitude modes from the simulation, and the overall pattern of the generated modes is well reproduced by the CNN. The CNN overpredicts the amplitudes of higher order modes, although these are two or more orders of magnitude smaller than the amplitude of the input mode. The CNN only has significant error for the smallest modes, which again demonstrates the resolution limit of the model. 

	In Fig. \ref{fig:pair_2lpt}, we show the amplitudes of the input mode along with the modes generated in 2LPT as a function of $k_2$. The green dash-dotted line shows the expected 2LPT amplitudes. The vertical dashed line shows the scale corresponding to the physical size of the CNN field of view. On scales larger than the field of view, the CNN underpredicts the N-body simulation amplitudes, which is not surprising since the CNN cannot fully model the interaction between modes on these scales. For scales where $k_2 > 0.2~\impc\, h$ the CNN overpredicts the amplitudes of the 2LPT modes. This is similar to the results for isolated waves, where the discreteness of the N-body evolution causes the small-scale modes to have lower amplitudes. Here again we see that the CNN predicts amplitudes closer to the continuum theory, since it assigns amplitudes close to the expected 2LPT values down to scales where $k_2\sim1~\impc\, h$. Notice that the presence of the second mode does not affect the predicted amplitude of the first mode, which is the same for all simulations.
	
	We find that the CNN is able to generalize well to these examples of coupled mode pairs, which are extremely different from the generic configurations of model's training data. The CNN model is able to accurately predict the 2LPT amplitudes over the range of scales where we expect it to perform well. This demonstrates that the network parameters encode information about the shapes of the mode coupling kernels in Eq. (\ref{eq:nlphi}--\ref{eq:nlA}). The errors of the CNN model reveal its limitations in terms of field of view and resolution, which can be used as diagnostics for future model improvement. 

	\section{Conclusion}

    We have trained CNNs to model the full phase space evolution of N-body simulations. The model has been designed to preserve the correct evolution on large, linear scales, and effectively nonlinearize the evolution on smaller scales. Our models are accurate to a few percent deep into the nonlinear regime, at small scales of $\sim 1\ \mpc\, h^{-1}$. This is a significant improvement in accuracy over other fast approximate N-body solvers such as COLA. We tested the robustness of our model in three idealized scenarios where either exact or perturbative theoretical calculations can be carried out. These include spherically symmetric configurations, isolated plane waves, and pairs of coupled plane waves. Each of these represents a dramatic departure from the random Gaussian initial conditions of the full N-body simulations, as in Eq. (\ref{eq:ics}), which were used in model training. In all of these scenarios the CNN model was able to accurately reproduce the correct evolution across a wide range of scales from the quasilinear down to the deeply nonlinear regime. Therefore, the model parameters have successfully encoded the dynamics of mode coupling represented in Eqs. (\ref{eq:nlphi}--\ref{eq:nlA}). The regimes where the CNN model made significant errors can be understood in terms of its limited field of view on large scales and its finite resolution on small scales. In particular, the model errors were most significant in regions where the matter distribution has strong gradients, such as at the edge of a spherical overdensity or at the nodes of an isolated plane wave. This demonstrates how these idealized scenarios can reveal both the robustness and the limitations of such complex CNN models, and indicates which features of the data should be targeted for more precise modeling. These tests help build confidence in our model, paving the way to applications in precision cosmological inference, where accelerating predictions of the nonlinear clustering is crucial for obtaining optimal parameter constraints.
    
	\section{Methods}
	\label{sec:meth}
	
    \subsection{\textbf{Modeling Structure Formation by Simulation or Perturbation Theory}}
	\label{ssec:meth:nb}
	
	Cosmological structure formation can only be accurately modeled by perturbation theories in the large-scale, perturbative regime, but requires N-body simulations in the deeply nonlinear, small-scale regime. Since our Universe begins with a nearly uniform, Gaussian initial condition, N-body simulations start with particles slightly perturbed from a uniform configuration, often a grid. The initial displacement of each particle $\mathbf{\Psi}$ from its grid location $\mathbf{q}$ can be accurately computed with LPT, as described in the previous section. The ZA predicts linear particle motion $\mathbf{\Psi}_\mathrm{ZA}(a,\mathbf{q}) = D(a) / D(1) \mathbf{\Psi}_\mathrm{ZA}(1, \mathbf{q})$, where $D(a)$ is the linear growth factor. Since only the growing mode is kept, the ZA velocity field is linearly related to the displacement field through $\mathbf{v}_\mathrm{ZA}(a, \mathbf{q}) = a H(a) f(a) \mathbf{\Psi}_\mathrm{ZA}(a, \mathbf{q})$, where $f(a) =\dee \log D(a)/\dee\log a$ is the linear growth rate.

	Nonlinear structure formation occurs at late times on small scales, where the ZA and higher order LPT fails to predict the particle motion. Modeling these trajectories accurately requires expensive N-body simulations, which typically integrate thousands of time steps. On the other hand, the Universe remains relatively uniform on the large scales where perturbation theories still apply. This motivates us to build our model to reproduce the small-scale N-body structure formation, while preserving the ZA evolution on large scales. Convolutional neural networks are well suited for this purpose, since they excel at local textures without a global view beyond their finite receptive fields. In other words, our goal is to train CNN models to nonlinearize the linear ZA inputs to reproduce the N-body simulation outputs.

	For N-body data, we use 210 simulations from the publicly available Quijote suite~\citep{Quijote}: 180 for training, 20 for validation, and 10 for testing. Each simulation has $512^3$ particles in a 1 $(\mathrm{Gpc}\, h^{-1})^3$ box, and has cosmological parameters: $\om=0.3175$, $\Omega_\mathrm{b} = 0.049$, $\Omega_\Lambda = 0.6825$, $h = 0.6711$, $\sigma_8 = 0.834$, and $n_\mathrm{s} = 0.9624$. We compare our model to the popular method COLA \citep[COmoving Lagrangian Acceleration,][]{Tassev2013}, as implemented in \texttt{L-PICOLA}~\citep{Howlett2015}. COLA evolves particles relative to their second order LPT (2LPT) trajectories, using only tens of time steps, making it much faster than a full N-body simulation. We run \texttt{L-PICOLA} with the same configurations as the full N-body simulations. We train and test our models at redshift $z=0$, corresponding to today.

    \subsection{\textbf{Designing Networks with Physical Considerations}}
    \label{ssec:meth:nn}

	As explained above, we want the CNN to maintain the ZA predictions ($\mathbf\Psi_\mathrm{ZA}$ and $\mathbf v_\mathrm{ZA}$ at the output time of the simulation snapshot) on large scales, which are still perturbative, to account for the nonlocal gravitational interactions beyond the local views of the CNN.  The CNN takes the linear inputs and forms accurate small-scale patterns. This is easily achieved by a global residual operation that adds the linear input to the network output, i.e., the CNN effectively predicts the difference between the N-body and ZA displacements.

	The global residual connection has other advantages too. One is that the Galilean symmetry can be approximately preserved by effective data augmentations. This is because, even though ZA and N-body evolution may be affected by large-scale shifts in displacements or velocities, i.e., by adding a global constant vector, the residuals are not. The other advantage is that the large-scale dependence on cosmological parameters is automatically accounted for by the ZA inputs, so the CNNs only need to model their small-scale modulations, as explained below. Therefore we separately train two networks, one for displacements and one for velocities, with linearly related inputs.

	By construction, CNNs preserve the translational symmetry if they are fully convolutional. However, most deep learning applications break this by the nonperiodic paddings commonly adopted in computer vision tasks. We fully preserve the translational symmetry using the N-body periodic boundary condition. In addition, our models approximately preserves the rotational symmetry of the cubic geometry by data augmentations that rotate and reflect the input and output fields according to all 48 variants as in \cite{He:2018ggn}.

    \subsection{\textbf{Network Architecture and Training}}
    \label{ssec:meth:at}

	Before feeding the data through the CNN models it must be prepared in an image-like format. This is straightforward due to the nearly uniform initial condition of the Universe. Both the ZA particle displacements and nonlinear N-body displacements are functions of the Lagrangian or initial positions $\mathbf{q}$, which form a uniform grid. Therefore both of them are a 3D image, with 3 channels being the Cartesian components of the displacement fields. We format the velocity fields likewise.

	We adopt a simple U-Net / V-Net~\citep{UNet, VNet} architecture similar to that in \cite{AlvesdeOliveira:2020yix}. It uses 3 levels of resolution connected in a ``V'' shape, comprising first 2 downsampling layers and then 2 upsampling layers by stride-2 $2^3$ convolutions and stride-\nicefrac12 $2^3$ transposed convolutions, respectively. The input, resampling layers, and output are connected by 2 $3^3$ convolutional blocks. A residual connection \citep[ResNet,][]{He2016}, here a $1^3$ convolution instead of identity pass, is added over each block, as in V-Net. A batch normalization layer follows every convolution layer except the first one and the last two, and a leaky ReLU activation (with negative slope 0.01) follows each batch normalization, as well as the first and the second to last convolutions. As the original ResNet, the last activation in each residual block applies after the addition. The inputs to each downsampling layer (on the left side of ``V'') are concatenated to the outputs of the upsampling layers (on the right side of ``V'') of the same resolution, at the top 3 resolution levels. All layers have 64 channels, except those after concatenations (128), and the input and the output (3). Unlike the original U-Net architecture, we add the input displacement or velocity fields directly to the output, so that the network only needs to learn the nonlinear residuals, as we have mentioned above.

	In the Lagrangian description, the simplest loss functions for training our networks minimizes the residuals in the displacements. \cite{AlvesdeOliveira:2020yix} and \cite{LiEtAl2021} have shown that combining that with a loss on density can greatly improve the model performance in the Eulerian space. We can compute the particle positions using their displacements, and then the Eulerian density distribution described by the overdensity $\delta(\mathbf{x}) \equiv n(\mathbf{x}) / \bar{n} - 1$. $n(\mathbf{x})$ is the particle number in grid cell at $\mathbf{x}$, and $\bar{n}$ is its mean value. We use the CIC scheme to interpolate the particles to the grid. For our displacement model, the total loss function is $\log L_\delta+\lambda \log L_{\Psi}$, where $L_\delta$ and $L_{\Psi}$ are the mean squared error (MSE) losses on $n(\mathbf{x})$ and $\mathbf{\Psi}(\mathbf{q})$, respectively. Adding the two losses in logarithm allows the optimizer to ignore their absolute magnitudes and trade between their relative changes. The hyperparameter $\lambda$ controls this trade-off, and we set $\lambda=3$ in the displacement network for faster training. 
	
	For the velocity network, we similarly use the particle momenta, $\dot{\mathbf{\Psi}}$ along with a CIC mesh construction of the Eulerian momentum field, $\mathbf{p}(a,\mathbf{x})$. We use the loss $\log L_{\dot{\Psi}} + \log L_{p}$. We refer to the former term as the momentum because all particles carry the same amount of mass, so their velocities represent mass weighted velocities, rather than the volume weighted velocities of the Eulerian fluid description. We do not multiply by the particle mass in practice though, since this only adds a constant to the loss.

	Training neural networks on 3D data can easily be GPU-memory-bound tasks. The entire Quijote simulation field ($512^3$) cannot be fed at once to the network, so smaller chunks of size $128^3$ are used. To preserve the translational equivariance, we periodically pad the input cubes by 20 voxels per side, but disable paddings in the convolutions. We train with a batch size of 16 distributed on 16 Nvidia V100 GPUs, and use the Adam optimizer~\citep{Kingma2014} with learning rate $0.0004$, and hyperparameters $\beta_1 = 0.9$, $\beta_2 = 0.999$. We reduce the learning by half when the loss does not improve for 3 epochs.


\section{Competing interests}

The authors declare no conflict of interest.

\section{Author contributions statement}

DJ, YL, SHe, \& FVN designed research;
DJ, YL, SHe, \& RAO performed research;
DJ, YL, SHe, \& FVN contributed new reagents or analytic tools;
DJ, YL, SHe, \& RAO analyzed data;
DJ, YL, SHe, RAO, FVN, SHo, \& DNS wrote the paper.

\section{Data Availability}

Our trained neural network parameters are available at
\url{github.com/dsjamieson/map2map_fid}.
We perform training and testing with \texttt{map2map}
(\url{github.com/eelregit/map2map}).
\texttt{map2map} is a framework for field-to-field neural-network
emulators, based on \texttt{PyTorch}~\citep{pytorch}.
It implements the aforementioned mechanisms to fully preserve
translational equivariance including the boundary conditions, and to do
rotational data augementation, for arbitrary dimensional data.
It is also designed for efficient data loading for training, to not
starve the GPUs of data due to their high performance.

\section{Acknowledgments}

The Flatiron Institute is supported by the Simons Foundation.


\bibliographystyle{abbrvnat}
\bibliography{neural_network_interpretation}

\end{document}